# Edge chirality determination of graphene by Raman spectroscopy

YuMeng You, ZhenHua Ni, Ting Yu, ZeXiang Shen[a]

Division of Physics and Applied Physics, School of Physical and Mathematical Sciences, Nanyang Technological University, Singapore, 637371

Raman imaging on the edges of single layer micromechanical cleavage graphene (MCG) was carried out. The intensity of disorder-induced Raman feature (D band at ~1350 cm$^{-1}$) was found to be correlated to the edge chirality: it is stronger at the armchair edge and weaker at the zigzag edge. This shows that Raman spectroscopy is a reliable and practical method to identify the chirality of graphene edge and to help in determination of the crystal orientation. The determination of graphene chirality is critically important for fundamental study as well as for applications.

[a] Author to whom correspondence should be addressed. Electronic mail: zexiang@ntu.edu.sg.

Graphene has attracted great attention not only because it is the ideal material to study the fundamental properties of 2D nanostructures,[1] but also for its potential applications in future electronic devices.[2] The exceptionally high crystallization and unique electronic properties make graphene a promising candidate for ultrahigh speed nanoelectronics.[3, 4] Graphene nanoribbons (GNR) have been receiving remarkable attention.[5-8] It was predicted that GNR with certain edge chirality would open the bandgap[5, 6, 9] and show distinguish magnetic,[7, 8] optical[10] and superconductive[11] properties. The bandgap opening of GNR has already been experimentally verified.[5, 12] All these peculiar properties are strongly dependent on the edge chirality (zigzag or armchair). Current technique to fabricate such well defined GNR is e-beam lithography from graphene sheet,[6, 13] in which the determination of the graphene crystal orientation and edge chirality is highly desired. Conventional methods like TEM, XRD and STM are either destructive, very time consuming or nearly impossible to locate such small regions of interest.[14, 15] The edge scattering would also prevent clear observation in some techniques, such as STM. The increasing interests in graphene demand a fast and non-destructive method to determine the chirality of edges and the crystal orientation of graphene sheet.

Cleavage describes the tendency to cleave along preferred planes of crystalline materials. Cleavage is related to the crystal structure and orientation,[16] and is also very useful in techniques for cutting semiconducting wafers and gem stones. As a perfect single crystalline structure, the MCG sheet is also expected to have similar cleavage behaviors[17]. After studying hundreds of MCG pieces, we found that the angles between MCG edges have an average value equaling to multiples of 30°. Fig. 1(a) shows the optical image of a typical

MCG sheet and the angles between the edges as an example. Fig. 1(b) shows the measurement of angles. It can be clearly seen that most of the angles are distributed around n × 30$^o$, where n is an integer between 0 and 6. Such a distribution suggests that the carbon atoms along the graphene edges have either zigzag or armchair structures. It can be easily shown in Fig. 1(c), for graphene with ideal edges, when the angle between two adjacent edges is 30$^o$, 90$^o$ or 150$^o$, these two edges should have different chirality; and for the angles of 60$^o$ and 120$^o$, the pairs of edges are of similar chirality (either zigzag or armchair). Recently, SEM and TEM results have actually shown that the edge of graphene is not ideally smooth.[18] For such edges, both kinds of chirality should exist. However, the majority of carbon atoms still have the same chirality as the smooth one. Therefore, the angle and chirality of edges should result from a microscopic averaging effect, as the majority of the carbon atoms along the edge are arranged in one kind of chirality different from the other kind. As a result, the chirality we mentioned in this work is the average from the statistics.

As one of the most commonly used techniques to characterize carbon related materials, Raman spectroscopy plays a very important role in acquiring information on the physical, chemical and even electronic properties of graphene and graphene based devices[19, 20]. In this work, we are able to determine the edge chirality, hence the crystal orientation of graphene using the difference in intensity of the disorder-induced D band on the different chiralities of edges (stronger in armchair edges and weaker in zigzag edges). This provides an easy and nondestructive method to identify the edge chirality of graphene, which would help to speed up the practical applications of graphene nanoelectronic devices, such as GNR.

The MCG sheets are prepared using the common micromechanical cleavage method[21] and deposited on a 300 nm SiO$_2$ / Si substrate which provides a good optical contrast.[22] The statistical study of the angles is done by measuring the angles between adjacent edges of single layer graphene (SLG) sheets from optical images. The Raman study was carried out using a WITec CRM200 confocal microscopy Raman system with a 100X objective lens (NA=0.95). The excitation source is a double frequency Nd:YAG laser (532 nm, CNI Lasers). Raman images were generated by scanning the sample with step size of 100 nm.

A piece of SLG is used as the sample. The number of layer has been determined by Raman spectroscopy as well as by the contrast method.[22, 23] There are three edges. The angle ($\theta_1$) between edge 1 and 2 is 30° and the angle ($\theta_2$) between edge 2 and 3 is 120°. Based on previous discussions, edge 1 and 2 should have different chiralities, with one zigzag and one armchair edge; and edge 2 and 3 have the same chirality. The main question here would be whether it is possible to determine the chirality of each edge. Here, Raman spectroscopy proves to be critically useful.

An SLG normally shows three major Raman bands as the G band around 1580 cm$^{-1}$, a very weak D band around 1350 cm$^{-1}$ and a 2D band around 2670 cm$^{-1}$,[24] as shown in Fig 2(d). The G band is related to the in-plane vibrational mode which has been studied in detail in all the graphitic materials. The appearance of the D band and 2D band is related to the double resonance Raman scattering process, which consists of several steps. For the D band, an electron-hole pair is excited. Then the electron (or hole) is inelastically scattered by a phonon, following the elastic scattering of the electron (or hole) by a defect. Finally, the excited electron and phonon recombine.[25] For the 2D band, elastic scattering of the electron (or hole) in the above process is replaced by the second phonon.[26]

On the other hand, the edge structure of graphene, like a defect, is also necessary for the double resonance condition.[27] Such edge structures of highly ordered pyrolytic graphite (HOPG) have been studied by Cancado et al[27, 28]. They found that the D band that appears at the armchair edge of HOPG is much stronger than that at the zigzag edge. Their theoretical study was based on the model of single layer graphene. After applying double resonance theory and considering the one-dimension character of the edge, they claimed that, the double resonance process can only be fulfilled at an armchair edge (stronger D band); while for a zigzag edge, the resonance process is forbidden (weaker or vanished D band). Hence, this idea is used to distinguish the chirality of the graphene edge and we focus mainly on the disorder-induced D band of SLG in this work.

Raman images constructed by the intensities of different Raman bands from a region containing edge 1 and 2 are shown in Figs. 2(a)-(c). The bright part in Fig. 2(a) corresponds to the appearance of the G band. The G band intensity is distributed uniformly over the whole graphene sheet, indicating the good quality of the sample. Figs. 2(b) and (c) show the D band intensity of graphene with laser polarization in the horizontal and vertical directions respectively. From these images, we can see that the D band only appears at the edges and shows a very strong polarization dependence. On the other hand, both the G band and 2D band are independent of the laser polarization (results not shown). We put the sample in this orientation to ensure that both edges (1 and 2) make the same angle with the laser polarization, $\pm 15^\circ$ to the horizontal polarization and $\pm 75^\circ$ to the vertical polarization.

Therefore, the stronger D band intensity at edge 1 compared to that at edge 2 is not due to the polarization effect, but related to the carbon atom arrangement at the edge, i.e. the chirality.[27] The spectra collected from different spots in both polarizations are shown in Fig. 2(d). All of the spectra are recorded under the same conditions. Spectra a and b are recorded at edges 1 and 2 respectively, with horizontal laser polarization, which is almost parallel to the two edges. It is obvious that edge 1 has a much stronger D band than edge 2. Since this is not due to the polarization effect, we can now identify edge 1 as armchair edge while edge 2 as zigzag edge. As mentioned before, the graphene edges are not perfectly smooth, so edge 1 should also contain some carbon atoms in zigzag arrangement, while edge 2 contains a small fraction of atoms in armchair arrangement. Do note that the armchair or zigzag arrangement mentioned here for edges 1 and 2 should be a result of the majority of carbon atoms along the edge arranged in one kind of chirality (either armchair or zigzag). This can be demonstrated in spectrum b of Fig. 2(d), where a weak D band can be observed. This suggests that, in edge 2, there is still a small portion of carbon atoms in armchair arrangement. On the other hand, spectra c and d in Fig. 2e are recorded at edges 1 and 2 respectively, using a vertical laser polarization. Both spectra hardly show any D band because of the polarization effect. Raman images of SLG edges with different angles are compared in Fig. 3. In the case of $30^o$ and $90^o$, edges show different D band contrast as they have different chiralities. On the other hand, in the case of $60^o$, a similar D band contrast is observed as the two edges have same chirality.

To rule out the possibility that the results of Raman imaging are caused by a difference in focusing or edge nonuniformity, we carried out a statistical analysis of the Raman intensities at each edge. For the SLG discussed in Fig. 2, it has three edges with an angle of $30^o$ between edge 1 and 2, and 120° between edge 2 and 3. To compare the three edges under the same conditions, Raman imaging is carried out on each individual edge, with the laser polarization parallel to that edge (Raman images not shown). The D band obtained at different spots of the edge was then fitted using a Lorentzian function and the intensity image was plotted. We manually chose the data points along the edges with similar area (~300 nm by 2 um) and calculated the average D band intensity generated from the edges. The average D band intensity from edge 1 (33.5) is obviously stronger than those from edges 2 (20.2) and 3 (22.0), which reveals that edge 1 is armchair and edges 2 and 3 are zigzag. This is consistent with the discussion in figure 2 about the chiralities of edges 1 and 2. Do note that even on the zigzag-like edge, the D band intensity is not zero, because the edges of MCG are not perfectly smooth. Similar results also were obtained for other MCG sheets. We have in total measured nine pairs of edges with different angles (listed in table 1), and our Raman results agree well with expectation. For angles of 30° and 90°, two adjacent edges show different D band intensities, indicating they have different atomic arrangements at the edges. While for angles of 60° and 90°, two adjacent edges show similar D band intensities, showing that they have the same arrangement. The intensity ratio for the same chirality edge is around 1.0, while the difference for different chirality edges is greater than 1.6, which suggests that Raman spectroscopy is a practical and reliable method for determination of the graphene edge structure. By knowing the edge arrangement, we can actually know the orientation of the whole graphene sheet. This is significant in the process of making graphene nano-constriction using lithography techniques[13, 29].

To conclude, we found that although the edges of MCG are not ideally smooth, they still show zigzag or armchair behaviors statistically. The defect-induced D band at the edges was found to be strongly polarization dependent, which is similar to that of graphite edges. This can be used to determine the chirality of the edges of the MCG. By applying this method, graphene nanoribbons fabricated by e-beam lithography can achieve better edge quality, thus helping to improve the performance of graphene based nanoelectronics.

Table 1. Statistic results of intensity of D band along different edges.

| Angle (degree) | # of pixels used for calculation | Average D band intensity (A. U.) | Standard deviation (A. U.) | Difference | Edge chirality |
|---|---|---|---|---|---|
| 30 | 63 | 20.2 | 6.8 | 1.66 | Zigzag |
| | 60 | 33.5 | 9.9 | | Armchair |
| 30 | 54 | 6.6 | 6.1 | 3.12 | Zigzag |
| | 75 | 20.6 | 6.3 | | Armchair |
| 90 | 42 | 8.1 | 4.4 | 3.40 | Zigzag |
| | 62 | 27.5 | 7.9 | | Armchair |
| 90 | 33 | 24.0 | 5.9 | 1.95 | Zigzag |
| | 34 | 46.8 | 7.4 | | Armchair |
| 60 | 107 | 13.9 | 6.3 | 1.01 | Same chirality |
| | 113 | 14.1 | 10.1 | | |
| 60 | 66 | 11.3 | 6.3 | 1.07 | Same chirality |
| | 66 | 12.1 | 6.4 | | |
| 120 | 63 | 20.2 | 6.8 | 1.09 | Same chirality |
| | 63 | 22.0 | 8.3 | | |
| 120 | 33 | 10.0 | 5.3 | 1.22 | Same chirality |
| | 30 | 12.2 | 4.7 | | |
| 120 | 37 | 33.9 | 8.2 | 1.10 | Same chirality |
| | 33 | 37.4 | 9.4 | | |

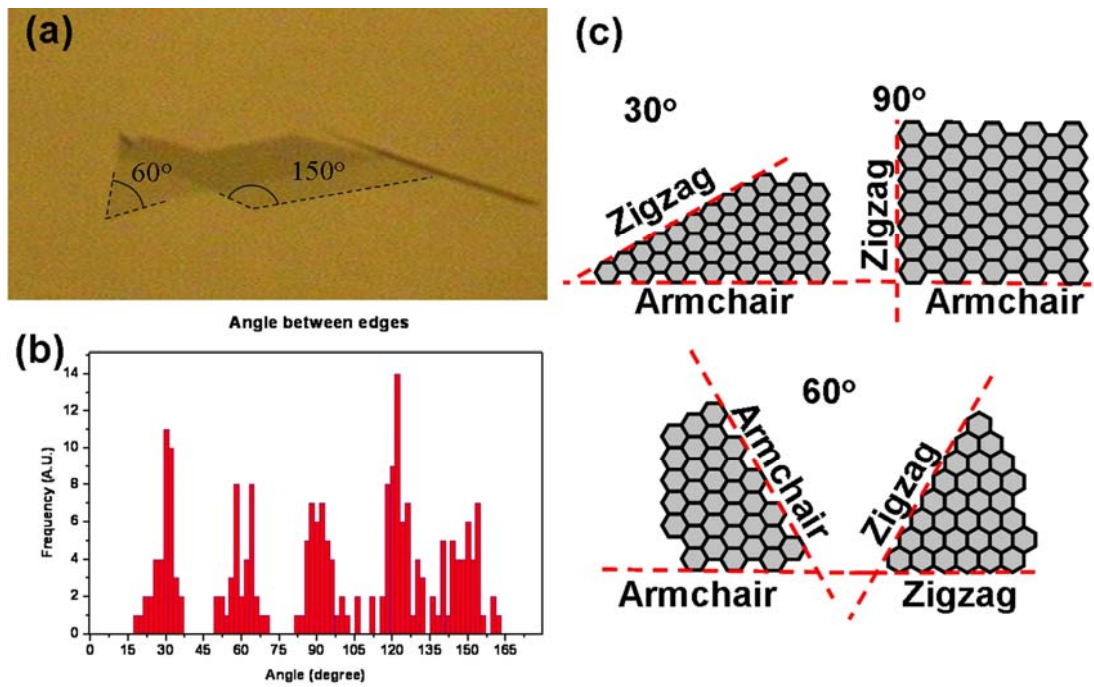

Fig. 1. (a) Optical image of a typical MCG sheet and the angles between edges. (b) The statistical results of the angle measurements. The standard deviation is 5.4$^\circ$. (c) Illustration of the relationship between angles and the chiralities of the adjacent edges.

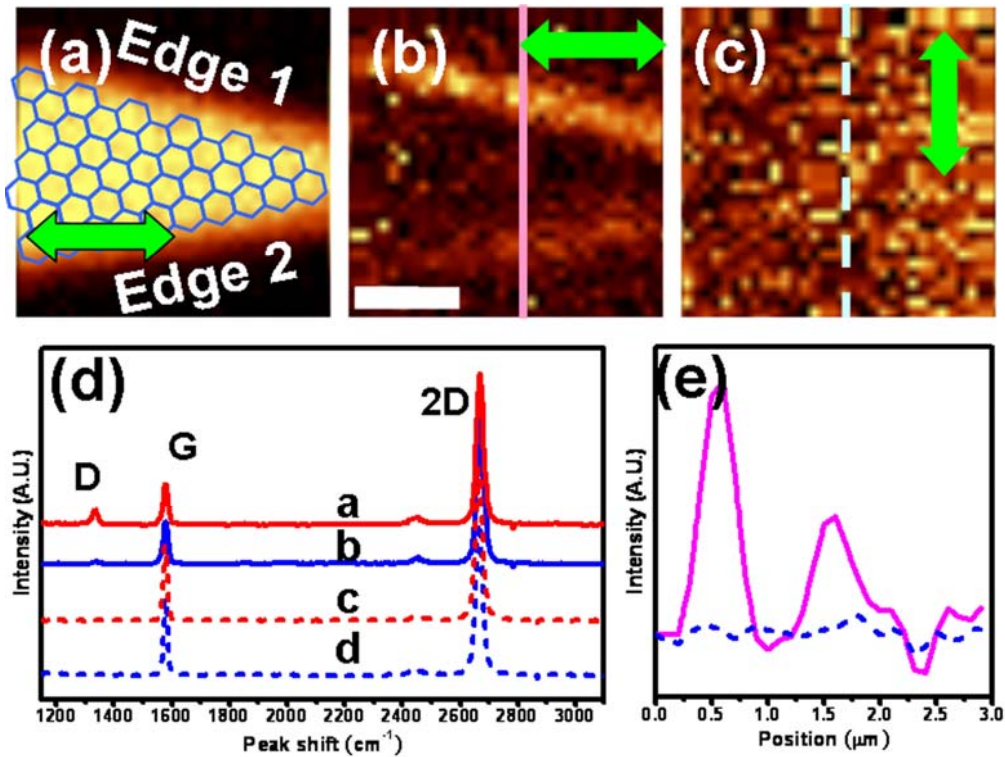

Fig. 2. (a) Raman image constructed by the intensity of G band with the expected arrangement in blue. Figs. (b) and (c) are images constructed by the D band intensity with horizontal and vertical polarization, respectively. All images share the same scale bar as indicated in Fig. (c) which is 2 μm. (d) Raman spectra taken from edge 1 (spectrum a), and edge 2 (spectrum b), with horizontal laser polarization. Spectra c and d were also collected from edges 1 and 2 respectively, with vertical laser polarization. (e) The solid and dotted lines represent the D band intensity profile (solid/dash) plotted along the solid line on Fig. 2(b) and the dashed line on Fig. (c), respectively.

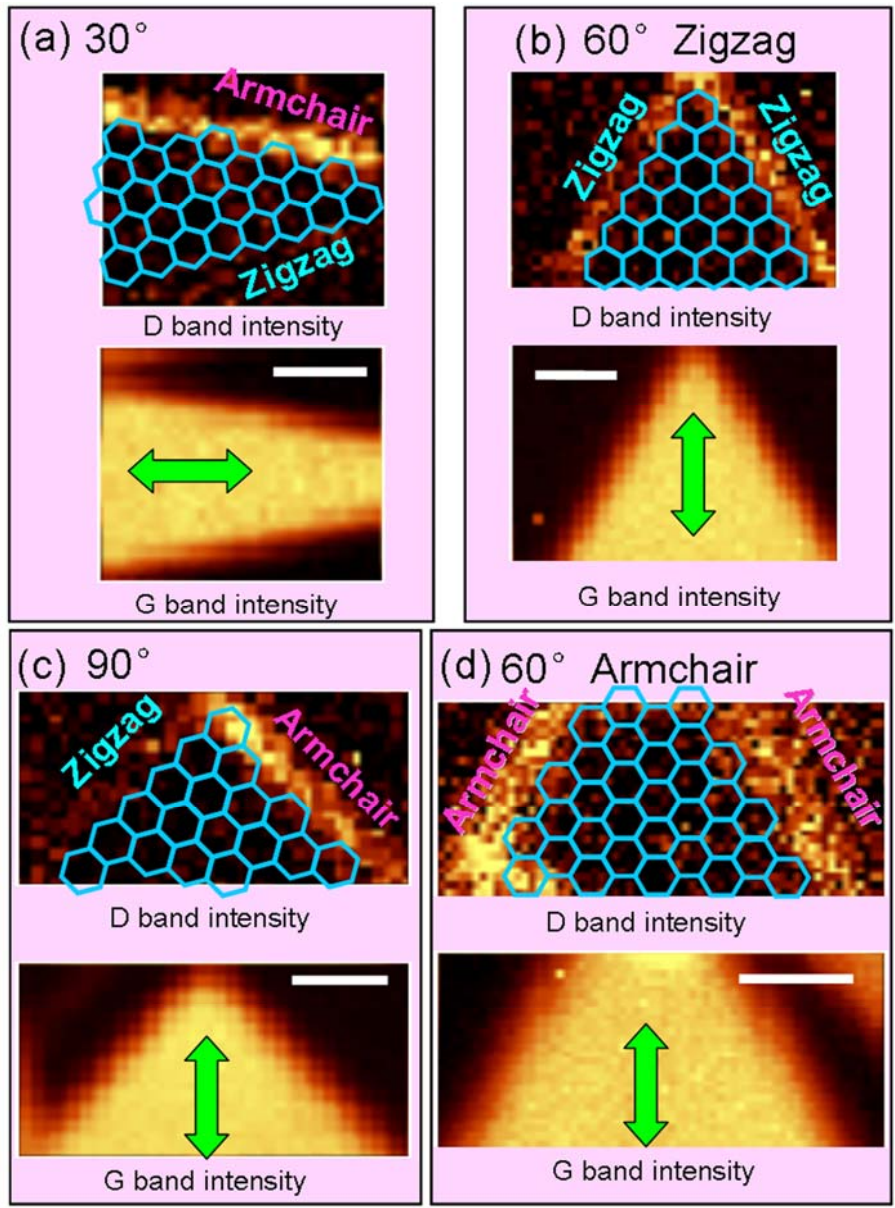

Fig. 3. Raman imaging results from edges with angles (a) 30$^o$, (b) 60$^o$ (zigzag), (c) 90$^o$ and (d) 60$^o$ (armchair). The images constructed by the G band intensity show the positions and shapes of the SLG sheets. The laser polarization is indicated by the green arrows. The superimposed frameworks are guides for the eye indicating the edge chirality. Note that the chirality of (b) and (d) were determined by the other pair of edges (not shown) with 30$^o$ / 90$^o$ on the same piece of SLG. The scale bar is 1 μm.